\documentclass[aip, rsi, graphicx, amsmath, amssymb, reprint]{revtex4-1}

\draft 

\usepackage{graphicx}

\begin{document}

\title{Entangled two-photon absorption for the continuous generation of excited state populations in plasma}

\author{David R. Smith}
\email{david.smith@wisc.edu}
\author{Matthias Beuting}
\author{Daniel J. Den Hartog}
\author{Benedikt Geiger}
\author{Scott T. Sanders}
\author{Xuting Yang}
\author{Jennifer T. Choy}
\affiliation{University of Wisconsin-Madison, Madison, WI}

\date{\today}

\begin{abstract}
Entangled two-photon absorption (ETPA) may be a viable technique to continuously drive an excited state population in plasma for high-bandwidth spectroscopy measurements of localized plasma turbulence or impurity density.
Classical two-photon absorption commonly requires a high-intensity, pulsed laser, but entangled photons with short entanglement time and high time correlation may allow for ETPA using a lower intensity, continuous-wave laser.
Notably, ETPA with non-collinear entangled photon generation allows for cross-beam spatial localization of the absorption or fluorescence signal using a single laser source.
Entangled photon generation, the ETPA cross-section, candidate transitions for an Ar-II species, and plans for a proof-of-principle measurement in a helicon plasma are discussed.
\end{abstract}

\maketitle

\section{\label{sec:intro}Introduction}

Plasma spectroscopy for magnetic fusion energy \cite{pawelec_core_2021} typically observes collisionally excited intrinsic impurities, injected impurities, or injected neutral beams via absorption or fluorescence collection. These measurements offer spatially resolved information on plasma dynamics and composition limited by the excitation volume and collection optics.
Laser-induced fluorescence (LIF), especially in combination with multi-photon processes, allows for more-localized excitation by driving an excited state population with a laser source and observing the associated fluorescence \cite{green_direct_2020, nikolic_measurements_2015, bieber_measurements_2011}.  
For instance, local neutral density measurements in plasma have been demonstrated with two-photon absorption laser-induced fluorescence (TALIF) \cite{dogariu_diagnostic_2022, steinberger_two-photon_2021, elliott_novel_2016, galante_two_2014, magee_two_2012}.  
For hydrogen TALIF, a $\lambda\!=\!205.2$\,nm source drives the $n\!=\!1\!\rightarrow\!3$ transition (102.6\,nm), and the $n\!=\!3\!\rightarrow\!2$ fluorescence (656.5\,nm) is observed.

\begin{figure}
    \centering
    \includegraphics[width=3.375in]{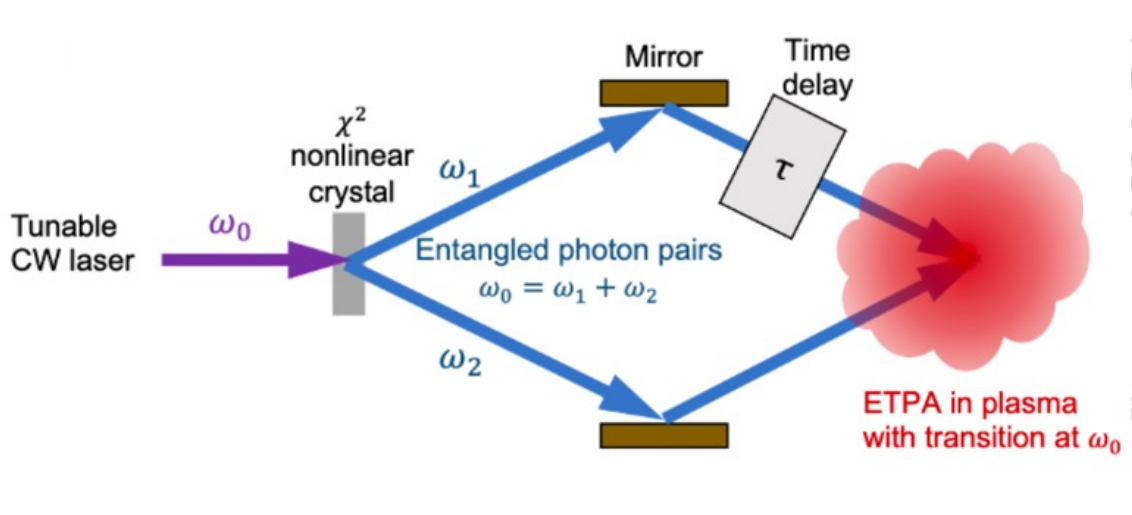}
    \caption{Illustration of an experimental configuration for ETPA measurements in plasmas.}
    \label{fig:ETPA_scheme}
\end{figure}

For multi-photon spectroscopy in plasmas, improved spatial resolution often comes at the expanse of poorer measurement signal-to-noise and bandwidth. The two-photon absorption (TPA) rate scales quadratically with photon flux for classical (coherent) light due to uncorrelated photon arrival, but the classical TPA cross-section is very small. 
Sufficient classical TPA rates require a pulsed laser (repetition rate $\sim$10\,Hz--100\,kHz) with high peak photon flux to leverage the quadratic scaling and overcome the small classical TPA cross-section.
However, continuous TPA with a narrow-linewidth CW (continuous-wave) source would be desirable for high bandwidth ($\sim$MHz) spectroscopy measurements of plasma turbulence and impurity density (Figure~\ref{fig:ETPA_scheme}), and that is the topic of this paper.

Entangled two-photon absorption (ETPA) \cite{schlawin_two-photon_2024, eshun_entangled_2022, raymer_how_2021, burdick_enhancing_2021} may be a viable technique to continuously drive \cite{villabona-monsalve_measurements_2020} an excited state population in plasma.
In contrast to classical light, entangled photon pairs are highly correlated which leads to the ETPA rate scaling linearly with photon flux. 
In addition, high time-correlation for entangled photons enhances the ETPA cross-section relative to the classical TPA cross-section. 
In this paper, we explore the feasibility to continuously drive an excited state population with ETPA and a CW laser to support high-bandwidth spectroscopy measurements of plasma turbulence and impurity density.  
Additional details of ETPA and entangled photon generation are given in Section~\ref{sec:etpa}. 
Section~\ref{sec:ar-ii} describes candidate two-photon transitions for an Ar-II (Ar+1) species in a $\sim$1--4\,eV laboratory plasma. 
Finally, Section~\ref{sec:conclusion} provides a summary and future directions.

\section{\label{sec:etpa}ETPA absorption rates and entangled photon generation}
\begin{figure}
    \centering
    \includegraphics[width=3.375in]{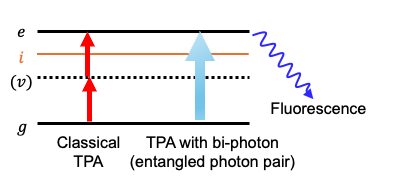}
    \caption{Energy-level diagram in a TPA process, in which two photons are involved in the excitation from a lower-energy ground state ($g$) to a higher-energy state ($e$). A short-lived, virtual state ($v$) is briefly populated in classical TPA, and this state may have neighboring real intermediate energy states ($i$) for which single-photon transitions are allowed. In ETPA, the involved photon pair (bi-photon) is strongly correlated.}
    \label{fig:tpa_energy}
\end{figure}

Classical TPA involves the excitation from a lower energy state to and excited state by sequential absorption of two (uncorrelated) photons through a virtual intermediate state (Figure~\ref{fig:tpa_energy})~\cite{eshun2022entangled}. The TPA rate per atom (units 1/s) for classical photons is $R=\sigma_c\,\phi^2$ where $\phi\!=\!I/hf$ is the photon flux, $I\!=\!\tfrac{1}{2}\,c\epsilon_0E^2$ is the field intensity, $\sigma_c$ is the classical TPA cross-section (units L$^4$T).
As previously mentioned, small $\sigma_c\!\approx\!10^{-48}$~cm$^4$s and the quadratic scaling with $\phi$ typically require a high intensity, short-pulse laser source.  
For example, an 1-W CW laser with a focal spot diameter of 10\,$\mu$m corresponds to a photon flux of about $10^{28}$\,photons/m$^2$/s, so the classical TPA rate per atom is $\sim\!10^8$\,1/s. 
In contrast, a pulsed laser with an equivalent (1 W) average power, 100\,mJ/pulse, 100\,ps pulse width, and 10\,Hz repetition rate generates the same average photon flux as the CW laser, but the peak power and peak photon flux are $10^9$ higher and the average classical TPA rate per atom is $\sim\!10^{17}$\,1/s.

ETPA utilizes an entangled photon pair (sometimes called a ``bi-photon'') in which the involved photons have strong correlations in their arrival time, momentum, energy, and polarization. Excitation from lower to higher energy states through ETPA can be considered as a single absorption event, leading to a linear scaling in the absorption rate with respect to incident light intensity~\cite{eshun2022entangled}.

The ETPA rate for entangled photons is~\cite{javanainen1990linear}
\begin{equation}
    R=\sigma_e\,\phi+\sigma_c\,\phi^2
    \label{eq:tpa_rate}
\end{equation}
where $\sigma_e$ is the ETPA cross-section (units L$^2$). At low entangled photon flux, the ETPA linear scaling with $\phi$ can dominate the classical TPA quadratic scaling with $\phi^2$.  Heuristically, the ETPA cross section is estimated to be
\begin{equation}
    \sigma_e\approx\frac{\sigma_c}{A_e\tau_e}
\end{equation}
where $A_e$ is the entangled area (beam cross section) and $\tau_e$ is the entangled photon correlation time (or entanglement time) \cite{schlawin_two-photon_2024, raymer_entangled_2021, parzuchowski_setting_2021, dayan_nonlinear_2005, joobeur_coherence_1996, javanainen_linear_1990}.
As described below, $\tau_e$ is the inverse of the large bandwidth of entangled photons, so $\tau_e$ is very short and $\sigma_e$ is strongly enhanced. Also, when the entangled photons converge at the target, small $A_e$ further enhances $\sigma_e$.

Entangled photon pairs with time-energy (frequency) entanglement can be created by spontaneous parametric down-conversion (SPDC) of a pump laser \cite{zhang_spontaneous_2021, schwaller_optimizing_2022, couteau_spontaneous_2018, avenhaus_experimental_2009}.
Energy conservation ensures that an entangled photon pair retains the pump beam energy, $f_p\!=\!f_1+f_2$ where $f_p$ is the pump photon frequency, and the pump beam linewidth, $\delta f_p\!=\!\delta(f_1+f_2)$.
However, the frequencies $f_1$ and $f_2$ of the entangled photons are generally broadband with large $\Delta f\!=\!f_2-f_1$ set by the optics. The short entanglement correlation time $\tau_e$ is set by the large bandwidth $\Delta f$:
\begin{equation}
    \tau_e\equiv\mbox{FFT}(\Delta f)=\Delta t=t_2-t_1.
\end{equation}
The simultaneous properties of short correlation time $\tau_e$ and narrow-band total energy $\delta(f_1+f_2)$ for entangled photon pairs may seem to violate a time-energy uncertainty limit, but the correlation time and total energy are decoupled (commute) for the entangled photon pair and are not subject to a joint uncertainty limit. 

As an example, consider a target species with a classical TPA cross-section $\sigma_c\!=\!10^{-48}$\,cm$^4$s. With an entanglement area $A_e\!=$\,(10$^{-3}$\,cm)$^2$ and a 10\,fs entanglement time (corresponding to a $\sim$20\,nm linewidth), the ETPA cross-section is $\sigma_e\approx10^{-28}$\,cm$^2$.

Note that correlation time and photon correlation refer to different statistical properties.  
The correlation time $\tau_e$ is the characteristic width of the second-order coherence $g^{(2)}(\tau)$ where $\tau=t_2-t_1$ is the delay time. The photon correlation is the second-order coherence at zero time delay, $g^{(2)}(0)$, and it possesses no upper bound.  
For coherent (classical) light with uncorrelated photons, $g^{(2)}(\tau)\!=\!1$ (the classical lower bound). 
For highly correlated entangled photons, $g^{(2)}(0)\!\gg\!1$.  
Short $\tau_e$ increases the ETPA cross-section $\sigma_e$, but the large correlation $g^{(2)}(0)\!\gg\!1$ leads to the ETPA rate scaling linearly with photon flux. 
Also, note that large photon correlation $g^{(2)}(0)\!\gg\!1$ is not restricted to entangled photon pairs nor is it inherently a quantum phenomenon. 
``Squeezed'' light with reduced phase fluctuations and enhanced field amplitude fluctuations also exhibits enhanced photon correlation and a TPA rate that scales linearly with photon flux \cite{georgiades_nonclassical_1995}.

Entangled photons generated by SPDC \cite{zhang_spontaneous_2021, schwaller_optimizing_2022, couteau_spontaneous_2018, avenhaus_experimental_2009} can be either collinear or non-collinear.  Also, the entangled photons have parallel polarization for type-I SPDC and perpendicular polarization for type-II SPDC.
Therefore, the ETPA configuration can either be along a single beam path or localized at a cross-beam intersection.  
Single-photon absorption or classical TPA along a single beam path can be susceptible to spurious reflections of fluorescing light that impair measurement localization and interpretation. 
Classical TPA with a single laser wavelength can drive TPA along the beam path with the potential for spurious reflections, but a sharp beam waist concentrates TPA at the waist \cite{bednar_assessment_2006}.
However, classical TPA with different wavelength lasers would allow for cross-beam localization such that any observed fluorescence would have originated at the cross-beam intersection.  
ETPA with non-collinear SPDC, on the other hand, allows for cross-beam localization with a single pump laser source, so again, any observed fluorescence would have originated at the cross-beam intersection. 
Finally, we note that demonstration of CW ETPA has been reported in chromophore solutions in Ref.\ \onlinecite{villabona-monsalve_measurements_2020}.
Entangled photons were produced at a rate of $10^7$\,1/s using a 70\,mW CW laser and collinear type-I SPDC.

\section{\label{sec:ar-ii}Two-photon transitions for Ar-II}

\begin{figure}
    \centering
    \includegraphics[width=3.375in]{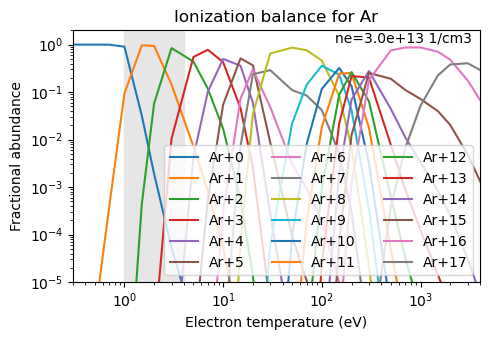}
    \caption{The fractional abundance of Ar charge states as a function of electron temperature for $n_e\!=\!3\!\times\!10^{13}$~1/cm$^3$. The grey box marks 1--4\,eV which is characteristic of a helicon plasma source.}
    \label{fig:ar_ion_balance}
\end{figure}

We consider Argon as a target species for ETPA in plasma because Ar charge states span $\sim$eV laboratory plasmas and $\sim$keV fusion plasmas, as shown in Figure~\ref{fig:ar_ion_balance}. 
Also, Ar is a common working gas in laboratory plasmas and a common injected impurity in fusion plasmas. 
The ionization balance in Figure~\ref{fig:ar_ion_balance} uses effective ionization and recombination data for ground state ions from the \texttt{ADAS} atomic data suite \cite{adas_2024}. 
We aim to demonstrate CW ETPA in a laboratory plasma, and the grey box in Figure~\ref{fig:ar_ion_balance} corresponds to 1--4\,eV for a helicon plasma source. 
Therefore, we will target two-photon transitions in Ar-II or Ar-III species which cover the temperature range for a $\sim$2--3\,eV helicon plasma.

\begin{figure}
    \centering
    \includegraphics[width=3.375in]{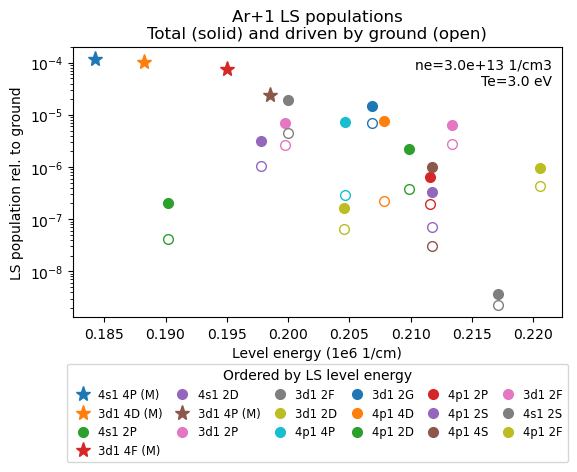}
    \caption{Collisional-radiative populations for LS-resolve energy levels for the Ar-II ion. Level energies are in $f/c$ wavenumber units. The stars denote metastable populations in addition to ground, the closed circles denote the level population driven by all metastables, and the open circles denote the level population driven by ground. Labels specify the valance electron configuration relative to {[}Ne{]}$3s^23p^4$ and the $^{2S+1}L$ term.}
    \label{fig:ar_ls_populations}
\end{figure}

A desirable two-photon transition would possess a large intrinsic population in the lower level and a small population in the upper level.  
Figure~\ref{fig:ar_ls_populations} shows intrinsic Ar-II populations for LS-resolved levels with electron density $3\times10^{13}$\,1/cm$^{3}$ and electron temperature 3\,eV for an \texttt{ADAS} collisional-radiative calculation. 
The calculation includes metastable transitions for the four largest populations in addition to the ground state. 
Another consideration is the wavelength of the two-photon transition.  
A pump laser wavelength below 400\,nm (above 3\,eV) would be desirable to reduce the contribution of electron impact excitation.  
Concurrently, a wavelength above 350\,nm would be desirable for compatibility with a frequency-doubled Ti:Sapphire laser (though higher order harmonics could be considered).  
A Ti:Sapphire laser is widely tunable over 700--1000\,nm, so the frequency doubled output can cover 350--500\,nm.
Therefore, we nominally target two-photon transitions compatible with an SPDC pump wavelength range of 350--400\,nm.

Finally, the candidate transitions should satisfy selection rules for two-phonton transitions.  
Ar is sufficiently low $Z$ that LS-coupling is the appropriate ordering for spin-orbit interactions (high $Z$ atoms and/or high charge states may require jj-coupling due to strong relativistic effects).  
Classical two-photon transitions satisfy the selection rules $\Delta S\!=\!0$ and $\Delta L\!=\!0,\pm2$, with the involved intermediate states obeying selection rules for single-photon transitions. We note that with the use of entangled photon pairs, time-frequency uncertainty can enable intermediate states that are higher energy~\cite{saleh1998entangled, burdick2018predicting}. 

\section{\label{sec:conclusion}Conclusion}

In conclusion, we suggest that ETPA may be a feasible technique to continuously drive an excited state population in plasma with a CW laser to support high-bandwidth spectroscopy measurements of plasma turbulence and impurity density.
Notably, ETPA with entangled photon generation by non-collinear SPDC allows for cross-beam spatial localization using a single laser source.
The ETPA rate scales linearly with photon flux, in contrast to the quadratic scaling for classical TPA, due to high time correlation in entangled photon pair arrival.
Also, the ETPA cross-section is enhanced by the short entanglement time.
Future work will attempt to demonstrate CW ETPA for an Ar-II or Ar-III two-photon transition in a helicon plasma source.
The target transition and entangled photon source will be developed based on entangled photon generation efficiency, ETPA cross-section and rate, intrinsic energy level populations, and fluorescence intensity.

\begin{acknowledgments}
DRS acknowledges helpful discussions with E.~Scime and T.~Steinberger. This material is based upon work supported by the U.S. Department of Energy, Office of Science, Office of Fusion Energy Sciences under Award Number DE-SC0024471. Data sharing is not applicable to this article as no new data were created or analyzed in this study.
\end{acknowledgments}

\bibliography{rsi_etpa,other}

\begin{thebibliography}{30}%
\makeatletter
\providecommand \@ifxundefined [1]{%
 \@ifx{#1\undefined}
}%
\providecommand \@ifnum [1]{%
 \ifnum #1\expandafter \@firstoftwo
 \else \expandafter \@secondoftwo
 \fi
}%
\providecommand \@ifx [1]{%
 \ifx #1\expandafter \@firstoftwo
 \else \expandafter \@secondoftwo
 \fi
}%
\providecommand \natexlab [1]{#1}%
\providecommand \enquote  [1]{``#1''}%
\providecommand \bibnamefont  [1]{#1}%
\providecommand \bibfnamefont [1]{#1}%
\providecommand \citenamefont [1]{#1}%
\providecommand \href@noop [0]{\@secondoftwo}%
\providecommand \href [0]{\begingroup \@sanitize@url \@href}%
\providecommand \@href[1]{\@@startlink{#1}\@@href}%
\providecommand \@@href[1]{\endgroup#1\@@endlink}%
\providecommand \@sanitize@url [0]{\catcode `\\12\catcode `\$12\catcode `\&12\catcode `\#12\catcode `\^12\catcode `\_12\catcode `\%12\relax}%
\providecommand \@@startlink[1]{}%
\providecommand \@@endlink[0]{}%
\providecommand \url  [0]{\begingroup\@sanitize@url \@url }%
\providecommand \@url [1]{\endgroup\@href {#1}{\urlprefix }}%
\providecommand \urlprefix  [0]{URL }%
\providecommand \Eprint [0]{\href }%
\providecommand \doibase [0]{http://dx.doi.org/}%
\providecommand \selectlanguage [0]{\@gobble}%
\providecommand \bibinfo  [0]{\@secondoftwo}%
\providecommand \bibfield  [0]{\@secondoftwo}%
\providecommand \translation [1]{[#1]}%
\providecommand \BibitemOpen [0]{}%
\providecommand \bibitemStop [0]{}%
\providecommand \bibitemNoStop [0]{.\EOS\space}%
\providecommand \EOS [0]{\spacefactor3000\relax}%
\providecommand \BibitemShut  [1]{\csname bibitem#1\endcsname}%
\let\auto@bib@innerbib\@empty
\bibitem [{\citenamefont {Pawelec}\ and\ \citenamefont {Contributors}(2021)}]{pawelec_core_2021}%
  \BibitemOpen
  \bibfield  {author} {\bibinfo {author} {\bibfnamefont {E.}~\bibnamefont {Pawelec}}\ and\ \bibinfo {author} {\bibfnamefont {J.~E.~T.}\ \bibnamefont {Contributors}},\ }\href {\doibase 10.1140/epjp/s13360-021-01800-6} {\bibfield  {journal} {\bibinfo  {journal} {Eur. Phys. J. Plus}\ }\textbf {\bibinfo {volume} {136}},\ \bibinfo {pages} {838} (\bibinfo {year} {2021})},\ \bibinfo {note} {number: 8 Publisher: Springer Berlin Heidelberg}\BibitemShut {NoStop}%
\bibitem [{\citenamefont {Green}, \citenamefont {Schmitz},\ and\ \citenamefont {Zepp}(2020)}]{green_direct_2020}%
  \BibitemOpen
  \bibfield  {author} {\bibinfo {author} {\bibfnamefont {J.}~\bibnamefont {Green}}, \bibinfo {author} {\bibfnamefont {O.}~\bibnamefont {Schmitz}}, \ and\ \bibinfo {author} {\bibfnamefont {M.}~\bibnamefont {Zepp}},\ }\href {\doibase 10.1063/1.5129232} {\bibfield  {journal} {\bibinfo  {journal} {Physics of Plasmas}\ }\textbf {\bibinfo {volume} {27}},\ \bibinfo {pages} {043511} (\bibinfo {year} {2020})},\ \bibinfo {note} {publisher: AIP Publishing LLC AIP Publishing}\BibitemShut {NoStop}%
\bibitem [{\citenamefont {Nikolić}\ \emph {et~al.}(2015)\citenamefont {Nikolić}, \citenamefont {Newton}, \citenamefont {Sukenik}, \citenamefont {Vušković},\ and\ \citenamefont {Popović}}]{nikolic_measurements_2015}%
  \BibitemOpen
  \bibfield  {author} {\bibinfo {author} {\bibfnamefont {M.}~\bibnamefont {Nikolić}}, \bibinfo {author} {\bibfnamefont {J.}~\bibnamefont {Newton}}, \bibinfo {author} {\bibfnamefont {C.~I.}\ \bibnamefont {Sukenik}}, \bibinfo {author} {\bibfnamefont {L.}~\bibnamefont {Vušković}}, \ and\ \bibinfo {author} {\bibfnamefont {S.}~\bibnamefont {Popović}},\ }\href {\doibase 10.1063/1.4905611} {\bibfield  {journal} {\bibinfo  {journal} {Journal of Applied Physics}\ }\textbf {\bibinfo {volume} {117}},\ \bibinfo {pages} {023304} (\bibinfo {year} {2015})},\ \bibinfo {note} {publisher: American Institute of Physics}\BibitemShut {NoStop}%
\bibitem [{\citenamefont {Bieber}\ \emph {et~al.}(2011)\citenamefont {Bieber}, \citenamefont {Bardin}, \citenamefont {Poucques}, \citenamefont {Brochard}, \citenamefont {Hugon}, \citenamefont {Vasseur},\ and\ \citenamefont {Bougdira}}]{bieber_measurements_2011}%
  \BibitemOpen
  \bibfield  {author} {\bibinfo {author} {\bibfnamefont {T.}~\bibnamefont {Bieber}}, \bibinfo {author} {\bibfnamefont {S.}~\bibnamefont {Bardin}}, \bibinfo {author} {\bibfnamefont {L.~d.}\ \bibnamefont {Poucques}}, \bibinfo {author} {\bibfnamefont {F.}~\bibnamefont {Brochard}}, \bibinfo {author} {\bibfnamefont {R.}~\bibnamefont {Hugon}}, \bibinfo {author} {\bibfnamefont {J.-L.}\ \bibnamefont {Vasseur}}, \ and\ \bibinfo {author} {\bibfnamefont {J.}~\bibnamefont {Bougdira}},\ }\href {\doibase 10.1088/0963-0252/20/1/015023} {\bibfield  {journal} {\bibinfo  {journal} {Plasma Sources Sci. Technol.}\ }\textbf {\bibinfo {volume} {20}},\ \bibinfo {pages} {015023} (\bibinfo {year} {2011})}\BibitemShut {NoStop}%
\bibitem [{\citenamefont {Dogariu}\ \emph {et~al.}(2022)\citenamefont {Dogariu}, \citenamefont {Cohen}, \citenamefont {Jandovitz}, \citenamefont {Vinoth}, \citenamefont {Evans},\ and\ \citenamefont {Swanson}}]{dogariu_diagnostic_2022}%
  \BibitemOpen
  \bibfield  {author} {\bibinfo {author} {\bibfnamefont {A.}~\bibnamefont {Dogariu}}, \bibinfo {author} {\bibfnamefont {S.~A.}\ \bibnamefont {Cohen}}, \bibinfo {author} {\bibfnamefont {P.}~\bibnamefont {Jandovitz}}, \bibinfo {author} {\bibfnamefont {S.}~\bibnamefont {Vinoth}}, \bibinfo {author} {\bibfnamefont {E.~S.}\ \bibnamefont {Evans}}, \ and\ \bibinfo {author} {\bibfnamefont {C.~P.~S.}\ \bibnamefont {Swanson}},\ }\href {\doibase 10.1063/5.0101683} {\bibfield  {journal} {\bibinfo  {journal} {Review of Scientific Instruments}\ }\textbf {\bibinfo {volume} {93}},\ \bibinfo {pages} {093519} (\bibinfo {year} {2022})}\BibitemShut {NoStop}%
\bibitem [{\citenamefont {Steinberger}\ \emph {et~al.}(2021)\citenamefont {Steinberger}, \citenamefont {McLaughlin}, \citenamefont {Biewer}, \citenamefont {Caneses},\ and\ \citenamefont {Scime}}]{steinberger_two-photon_2021}%
  \BibitemOpen
  \bibfield  {author} {\bibinfo {author} {\bibfnamefont {T.~E.}\ \bibnamefont {Steinberger}}, \bibinfo {author} {\bibfnamefont {J.~W.}\ \bibnamefont {McLaughlin}}, \bibinfo {author} {\bibfnamefont {T.~M.}\ \bibnamefont {Biewer}}, \bibinfo {author} {\bibfnamefont {J.~F.}\ \bibnamefont {Caneses}}, \ and\ \bibinfo {author} {\bibfnamefont {E.~E.}\ \bibnamefont {Scime}},\ }\href {\doibase 10.1063/5.0054734} {\bibfield  {journal} {\bibinfo  {journal} {Physics of Plasmas}\ }\textbf {\bibinfo {volume} {28}},\ \bibinfo {pages} {082501} (\bibinfo {year} {2021})}\BibitemShut {NoStop}%
\bibitem [{\citenamefont {Elliott}, \citenamefont {Scime},\ and\ \citenamefont {Short}(2016)}]{elliott_novel_2016}%
  \BibitemOpen
  \bibfield  {author} {\bibinfo {author} {\bibfnamefont {D.}~\bibnamefont {Elliott}}, \bibinfo {author} {\bibfnamefont {E.}~\bibnamefont {Scime}}, \ and\ \bibinfo {author} {\bibfnamefont {Z.}~\bibnamefont {Short}},\ }\href {\doibase 10.1063/1.4955489} {\bibfield  {journal} {\bibinfo  {journal} {Review of Scientific Instruments}\ }\textbf {\bibinfo {volume} {87}},\ \bibinfo {pages} {11E504} (\bibinfo {year} {2016})}\BibitemShut {NoStop}%
\bibitem [{\citenamefont {Galante}, \citenamefont {Magee},\ and\ \citenamefont {Scime}(2014)}]{galante_two_2014}%
  \BibitemOpen
  \bibfield  {author} {\bibinfo {author} {\bibfnamefont {M.~E.}\ \bibnamefont {Galante}}, \bibinfo {author} {\bibfnamefont {R.~M.}\ \bibnamefont {Magee}}, \ and\ \bibinfo {author} {\bibfnamefont {E.~E.}\ \bibnamefont {Scime}},\ }\href {\doibase 10.1063/1.4873900} {\bibfield  {journal} {\bibinfo  {journal} {Physics of Plasmas}\ }\textbf {\bibinfo {volume} {21}},\ \bibinfo {pages} {055704} (\bibinfo {year} {2014})}\BibitemShut {NoStop}%
\bibitem [{\citenamefont {Magee}\ \emph {et~al.}(2012)\citenamefont {Magee}, \citenamefont {Galante}, \citenamefont {McCarren}, \citenamefont {Scime}, \citenamefont {Boivin}, \citenamefont {Brooks}, \citenamefont {Groebner}, \citenamefont {Hill},\ and\ \citenamefont {Porter}}]{magee_two_2012}%
  \BibitemOpen
  \bibfield  {author} {\bibinfo {author} {\bibfnamefont {R.~M.}\ \bibnamefont {Magee}}, \bibinfo {author} {\bibfnamefont {M.~E.}\ \bibnamefont {Galante}}, \bibinfo {author} {\bibfnamefont {D.}~\bibnamefont {McCarren}}, \bibinfo {author} {\bibfnamefont {E.~E.}\ \bibnamefont {Scime}}, \bibinfo {author} {\bibfnamefont {R.~L.}\ \bibnamefont {Boivin}}, \bibinfo {author} {\bibfnamefont {N.~H.}\ \bibnamefont {Brooks}}, \bibinfo {author} {\bibfnamefont {R.~J.}\ \bibnamefont {Groebner}}, \bibinfo {author} {\bibfnamefont {D.~N.}\ \bibnamefont {Hill}}, \ and\ \bibinfo {author} {\bibfnamefont {G.~D.}\ \bibnamefont {Porter}},\ }\href {\doibase 10.1063/1.4728092} {\bibfield  {journal} {\bibinfo  {journal} {Review of Scientific Instruments}\ }\textbf {\bibinfo {volume} {83}},\ \bibinfo {pages} {10D701} (\bibinfo {year} {2012})},\ \bibinfo {note} {publisher: American Institute of Physics}\BibitemShut {NoStop}%
\bibitem [{\citenamefont {Schlawin}(2024)}]{schlawin_two-photon_2024}%
  \BibitemOpen
  \bibfield  {author} {\bibinfo {author} {\bibfnamefont {F.}~\bibnamefont {Schlawin}},\ }\href {\doibase 10.1063/5.0196817} {\bibfield  {journal} {\bibinfo  {journal} {The Journal of Chemical Physics}\ }\textbf {\bibinfo {volume} {160}},\ \bibinfo {pages} {144117} (\bibinfo {year} {2024})}\BibitemShut {NoStop}%
\bibitem [{\citenamefont {Eshun}\ \emph {et~al.}(2022{\natexlab{a}})\citenamefont {Eshun}, \citenamefont {Varnavski}, \citenamefont {Villabona-Monsalve}, \citenamefont {Burdick},\ and\ \citenamefont {Goodson}}]{eshun_entangled_2022}%
  \BibitemOpen
  \bibfield  {author} {\bibinfo {author} {\bibfnamefont {A.}~\bibnamefont {Eshun}}, \bibinfo {author} {\bibfnamefont {O.}~\bibnamefont {Varnavski}}, \bibinfo {author} {\bibfnamefont {J.~P.}\ \bibnamefont {Villabona-Monsalve}}, \bibinfo {author} {\bibfnamefont {R.~K.}\ \bibnamefont {Burdick}}, \ and\ \bibinfo {author} {\bibfnamefont {T.~I.}\ \bibnamefont {Goodson}},\ }\href {\doibase 10.1021/acs.accounts.1c00687} {\bibfield  {journal} {\bibinfo  {journal} {Acc. Chem. Res.}\ }\textbf {\bibinfo {volume} {55}},\ \bibinfo {pages} {991} (\bibinfo {year} {2022}{\natexlab{a}})},\ \bibinfo {note} {publisher: American Chemical Society}\BibitemShut {NoStop}%
\bibitem [{\citenamefont {Raymer}\ \emph {et~al.}(2021)\citenamefont {Raymer}, \citenamefont {Landes}, \citenamefont {Allgaier}, \citenamefont {Merkouche}, \citenamefont {Smith},\ and\ \citenamefont {Marcus}}]{raymer_how_2021}%
  \BibitemOpen
  \bibfield  {author} {\bibinfo {author} {\bibfnamefont {M.~G.}\ \bibnamefont {Raymer}}, \bibinfo {author} {\bibfnamefont {T.}~\bibnamefont {Landes}}, \bibinfo {author} {\bibfnamefont {M.}~\bibnamefont {Allgaier}}, \bibinfo {author} {\bibfnamefont {S.}~\bibnamefont {Merkouche}}, \bibinfo {author} {\bibfnamefont {B.~J.}\ \bibnamefont {Smith}}, \ and\ \bibinfo {author} {\bibfnamefont {A.~H.}\ \bibnamefont {Marcus}},\ }\href {\doibase 10.1364/OPTICA.426674} {\bibfield  {journal} {\bibinfo  {journal} {Optica}\ }\textbf {\bibinfo {volume} {8}},\ \bibinfo {pages} {757} (\bibinfo {year} {2021})}\BibitemShut {NoStop}%
\bibitem [{\citenamefont {Burdick}, \citenamefont {Schatz},\ and\ \citenamefont {Goodson}(2021)}]{burdick_enhancing_2021}%
  \BibitemOpen
  \bibfield  {author} {\bibinfo {author} {\bibfnamefont {R.~K.}\ \bibnamefont {Burdick}}, \bibinfo {author} {\bibfnamefont {G.~C.}\ \bibnamefont {Schatz}}, \ and\ \bibinfo {author} {\bibfnamefont {T.}~\bibnamefont {Goodson}},\ }\href {\doibase 10.1021/jacs.1c09728} {\bibfield  {journal} {\bibinfo  {journal} {J. Am. Chem. Soc.}\ }\textbf {\bibinfo {volume} {143}},\ \bibinfo {pages} {16930} (\bibinfo {year} {2021})}\BibitemShut {NoStop}%
\bibitem [{\citenamefont {Villabona-Monsalve}, \citenamefont {Burdick},\ and\ \citenamefont {Goodson}(2020)}]{villabona-monsalve_measurements_2020}%
  \BibitemOpen
  \bibfield  {author} {\bibinfo {author} {\bibfnamefont {J.~P.}\ \bibnamefont {Villabona-Monsalve}}, \bibinfo {author} {\bibfnamefont {R.~K.}\ \bibnamefont {Burdick}}, \ and\ \bibinfo {author} {\bibfnamefont {T.~I.}\ \bibnamefont {Goodson}},\ }\href {\doibase 10.1021/acs.jpcc.0c08678} {\bibfield  {journal} {\bibinfo  {journal} {J. Phys. Chem. C}\ }\textbf {\bibinfo {volume} {124}},\ \bibinfo {pages} {24526} (\bibinfo {year} {2020})},\ \bibinfo {note} {publisher: American Chemical Society}\BibitemShut {NoStop}%
\bibitem [{\citenamefont {Eshun}\ \emph {et~al.}(2022{\natexlab{b}})\citenamefont {Eshun}, \citenamefont {Varnavski}, \citenamefont {Villabona-Monsalve}, \citenamefont {Burdick},\ and\ \citenamefont {Goodson~III}}]{eshun2022entangled}%
  \BibitemOpen
  \bibfield  {author} {\bibinfo {author} {\bibfnamefont {A.}~\bibnamefont {Eshun}}, \bibinfo {author} {\bibfnamefont {O.}~\bibnamefont {Varnavski}}, \bibinfo {author} {\bibfnamefont {J.~P.}\ \bibnamefont {Villabona-Monsalve}}, \bibinfo {author} {\bibfnamefont {R.~K.}\ \bibnamefont {Burdick}}, \ and\ \bibinfo {author} {\bibfnamefont {T.}~\bibnamefont {Goodson~III}},\ }\href@noop {} {\bibfield  {journal} {\bibinfo  {journal} {Accounts of Chemical Research}\ }\textbf {\bibinfo {volume} {55}},\ \bibinfo {pages} {991} (\bibinfo {year} {2022}{\natexlab{b}})}\BibitemShut {NoStop}%
\bibitem [{\citenamefont {Javanainen}\ and\ \citenamefont {Gould}(1990{\natexlab{a}})}]{javanainen1990linear}%
  \BibitemOpen
  \bibfield  {author} {\bibinfo {author} {\bibfnamefont {J.}~\bibnamefont {Javanainen}}\ and\ \bibinfo {author} {\bibfnamefont {P.~L.}\ \bibnamefont {Gould}},\ }\href@noop {} {\bibfield  {journal} {\bibinfo  {journal} {Physical Review A}\ }\textbf {\bibinfo {volume} {41}},\ \bibinfo {pages} {5088} (\bibinfo {year} {1990}{\natexlab{a}})}\BibitemShut {NoStop}%
\bibitem [{\citenamefont {Raymer}, \citenamefont {Landes},\ and\ \citenamefont {Marcus}(2021)}]{raymer_entangled_2021}%
  \BibitemOpen
  \bibfield  {author} {\bibinfo {author} {\bibfnamefont {M.~G.}\ \bibnamefont {Raymer}}, \bibinfo {author} {\bibfnamefont {T.}~\bibnamefont {Landes}}, \ and\ \bibinfo {author} {\bibfnamefont {A.~H.}\ \bibnamefont {Marcus}},\ }\href {\doibase 10.1063/5.0049338} {\bibfield  {journal} {\bibinfo  {journal} {J. Chem. Phys.}\ }\textbf {\bibinfo {volume} {155}},\ \bibinfo {pages} {081501} (\bibinfo {year} {2021})}\BibitemShut {NoStop}%
\bibitem [{\citenamefont {Parzuchowski}\ \emph {et~al.}(2021)\citenamefont {Parzuchowski}, \citenamefont {Mikhaylov}, \citenamefont {Mazurek}, \citenamefont {Wilson}, \citenamefont {Lum}, \citenamefont {Gerrits}, \citenamefont {Camp}, \citenamefont {Stevens},\ and\ \citenamefont {Jimenez}}]{parzuchowski_setting_2021}%
  \BibitemOpen
  \bibfield  {author} {\bibinfo {author} {\bibfnamefont {K.~M.}\ \bibnamefont {Parzuchowski}}, \bibinfo {author} {\bibfnamefont {A.}~\bibnamefont {Mikhaylov}}, \bibinfo {author} {\bibfnamefont {M.~D.}\ \bibnamefont {Mazurek}}, \bibinfo {author} {\bibfnamefont {R.~N.}\ \bibnamefont {Wilson}}, \bibinfo {author} {\bibfnamefont {D.~J.}\ \bibnamefont {Lum}}, \bibinfo {author} {\bibfnamefont {T.}~\bibnamefont {Gerrits}}, \bibinfo {author} {\bibfnamefont {C.~H.}\ \bibnamefont {Camp}}, \bibinfo {author} {\bibfnamefont {M.~J.}\ \bibnamefont {Stevens}}, \ and\ \bibinfo {author} {\bibfnamefont {R.}~\bibnamefont {Jimenez}},\ }\href {\doibase 10.1103/PhysRevApplied.15.044012} {\bibfield  {journal} {\bibinfo  {journal} {Phys. Rev. Appl.}\ }\textbf {\bibinfo {volume} {15}},\ \bibinfo {pages} {044012} (\bibinfo {year} {2021})},\ \bibinfo {note} {publisher: American Physical Society}\BibitemShut {NoStop}%
\bibitem [{\citenamefont {Dayan}\ \emph {et~al.}(2005)\citenamefont {Dayan}, \citenamefont {Pe’er}, \citenamefont {Friesem},\ and\ \citenamefont {Silberberg}}]{dayan_nonlinear_2005}%
  \BibitemOpen
  \bibfield  {author} {\bibinfo {author} {\bibfnamefont {B.}~\bibnamefont {Dayan}}, \bibinfo {author} {\bibfnamefont {A.}~\bibnamefont {Pe’er}}, \bibinfo {author} {\bibfnamefont {A.~A.}\ \bibnamefont {Friesem}}, \ and\ \bibinfo {author} {\bibfnamefont {Y.}~\bibnamefont {Silberberg}},\ }\href {\doibase 10.1103/PhysRevLett.94.043602} {\bibfield  {journal} {\bibinfo  {journal} {Phys. Rev. Lett.}\ }\textbf {\bibinfo {volume} {94}},\ \bibinfo {pages} {043602} (\bibinfo {year} {2005})}\BibitemShut {NoStop}%
\bibitem [{\citenamefont {Joobeur}\ \emph {et~al.}(1996)\citenamefont {Joobeur}, \citenamefont {Saleh}, \citenamefont {Larchuk},\ and\ \citenamefont {Teich}}]{joobeur_coherence_1996}%
  \BibitemOpen
  \bibfield  {author} {\bibinfo {author} {\bibfnamefont {A.}~\bibnamefont {Joobeur}}, \bibinfo {author} {\bibfnamefont {B.~E.~A.}\ \bibnamefont {Saleh}}, \bibinfo {author} {\bibfnamefont {T.~S.}\ \bibnamefont {Larchuk}}, \ and\ \bibinfo {author} {\bibfnamefont {M.~C.}\ \bibnamefont {Teich}},\ }\href {\doibase 10.1103/PhysRevA.53.4360} {\bibfield  {journal} {\bibinfo  {journal} {Phys. Rev. A}\ }\textbf {\bibinfo {volume} {53}},\ \bibinfo {pages} {4360} (\bibinfo {year} {1996})}\BibitemShut {NoStop}%
\bibitem [{\citenamefont {Javanainen}\ and\ \citenamefont {Gould}(1990{\natexlab{b}})}]{javanainen_linear_1990}%
  \BibitemOpen
  \bibfield  {author} {\bibinfo {author} {\bibfnamefont {J.}~\bibnamefont {Javanainen}}\ and\ \bibinfo {author} {\bibfnamefont {P.~L.}\ \bibnamefont {Gould}},\ }\href {\doibase 10.1103/PhysRevA.41.5088} {\bibfield  {journal} {\bibinfo  {journal} {Phys. Rev. A}\ }\textbf {\bibinfo {volume} {41}},\ \bibinfo {pages} {5088} (\bibinfo {year} {1990}{\natexlab{b}})}\BibitemShut {NoStop}%
\bibitem [{\citenamefont {Zhang}\ \emph {et~al.}(2021)\citenamefont {Zhang}, \citenamefont {Huang}, \citenamefont {Liu}, \citenamefont {Li},\ and\ \citenamefont {Guo}}]{zhang_spontaneous_2021}%
  \BibitemOpen
  \bibfield  {author} {\bibinfo {author} {\bibfnamefont {C.}~\bibnamefont {Zhang}}, \bibinfo {author} {\bibfnamefont {Y.-F.}\ \bibnamefont {Huang}}, \bibinfo {author} {\bibfnamefont {B.-H.}\ \bibnamefont {Liu}}, \bibinfo {author} {\bibfnamefont {C.-F.}\ \bibnamefont {Li}}, \ and\ \bibinfo {author} {\bibfnamefont {G.-C.}\ \bibnamefont {Guo}},\ }\href {\doibase 10.1002/qute.202000132} {\bibfield  {journal} {\bibinfo  {journal} {Advanced Quantum Technologies}\ }\textbf {\bibinfo {volume} {4}},\ \bibinfo {pages} {2000132} (\bibinfo {year} {2021})},\ \bibinfo {note} {\_eprint: https://onlinelibrary.wiley.com/doi/pdf/10.1002/qute.202000132}\BibitemShut {NoStop}%
\bibitem [{\citenamefont {Schwaller}\ \emph {et~al.}(2022)\citenamefont {Schwaller}, \citenamefont {Park}, \citenamefont {Okamoto},\ and\ \citenamefont {Takeuchi}}]{schwaller_optimizing_2022}%
  \BibitemOpen
  \bibfield  {author} {\bibinfo {author} {\bibfnamefont {N.}~\bibnamefont {Schwaller}}, \bibinfo {author} {\bibfnamefont {G.}~\bibnamefont {Park}}, \bibinfo {author} {\bibfnamefont {R.}~\bibnamefont {Okamoto}}, \ and\ \bibinfo {author} {\bibfnamefont {S.}~\bibnamefont {Takeuchi}},\ }\href {\doibase 10.1103/PhysRevA.106.043719} {\bibfield  {journal} {\bibinfo  {journal} {Phys. Rev. A}\ }\textbf {\bibinfo {volume} {106}},\ \bibinfo {pages} {043719} (\bibinfo {year} {2022})},\ \bibinfo {note} {publisher: American Physical Society}\BibitemShut {NoStop}%
\bibitem [{\citenamefont {Couteau}(2018)}]{couteau_spontaneous_2018}%
  \BibitemOpen
  \bibfield  {author} {\bibinfo {author} {\bibfnamefont {C.}~\bibnamefont {Couteau}},\ }\href {\doibase 10.1080/00107514.2018.1488463} {\bibfield  {journal} {\bibinfo  {journal} {Contemporary Physics}\ }\textbf {\bibinfo {volume} {59}},\ \bibinfo {pages} {291} (\bibinfo {year} {2018})},\ \bibinfo {note} {arXiv:1809.00127 [physics, physics:quant-ph]}\BibitemShut {NoStop}%
\bibitem [{\citenamefont {Avenhaus}\ \emph {et~al.}(2009)\citenamefont {Avenhaus}, \citenamefont {Chekhova}, \citenamefont {Krivitsky}, \citenamefont {Leuchs},\ and\ \citenamefont {Silberhorn}}]{avenhaus_experimental_2009}%
  \BibitemOpen
  \bibfield  {author} {\bibinfo {author} {\bibfnamefont {M.}~\bibnamefont {Avenhaus}}, \bibinfo {author} {\bibfnamefont {M.~V.}\ \bibnamefont {Chekhova}}, \bibinfo {author} {\bibfnamefont {L.~A.}\ \bibnamefont {Krivitsky}}, \bibinfo {author} {\bibfnamefont {G.}~\bibnamefont {Leuchs}}, \ and\ \bibinfo {author} {\bibfnamefont {C.}~\bibnamefont {Silberhorn}},\ }\href {\doibase 10.1103/PhysRevA.79.043836} {\bibfield  {journal} {\bibinfo  {journal} {Phys. Rev. A}\ }\textbf {\bibinfo {volume} {79}},\ \bibinfo {pages} {043836} (\bibinfo {year} {2009})},\ \bibinfo {note} {publisher: American Physical Society}\BibitemShut {NoStop}%
\bibitem [{\citenamefont {Georgiades}\ \emph {et~al.}(1995)\citenamefont {Georgiades}, \citenamefont {Polzik}, \citenamefont {Edamatsu}, \citenamefont {Kimble},\ and\ \citenamefont {Parkins}}]{georgiades_nonclassical_1995}%
  \BibitemOpen
  \bibfield  {author} {\bibinfo {author} {\bibfnamefont {N.~P.}\ \bibnamefont {Georgiades}}, \bibinfo {author} {\bibfnamefont {E.~S.}\ \bibnamefont {Polzik}}, \bibinfo {author} {\bibfnamefont {K.}~\bibnamefont {Edamatsu}}, \bibinfo {author} {\bibfnamefont {H.~J.}\ \bibnamefont {Kimble}}, \ and\ \bibinfo {author} {\bibfnamefont {A.~S.}\ \bibnamefont {Parkins}},\ }\href {\doibase 10.1103/PhysRevLett.75.3426} {\bibfield  {journal} {\bibinfo  {journal} {Phys. Rev. Lett.}\ }\textbf {\bibinfo {volume} {75}},\ \bibinfo {pages} {3426} (\bibinfo {year} {1995})}\BibitemShut {NoStop}%
\bibitem [{\citenamefont {Bednar}, \citenamefont {Walewski},\ and\ \citenamefont {Sanders}(2006)}]{bednar_assessment_2006}%
  \BibitemOpen
  \bibfield  {author} {\bibinfo {author} {\bibfnamefont {N.~J.}\ \bibnamefont {Bednar}}, \bibinfo {author} {\bibfnamefont {J.~W.}\ \bibnamefont {Walewski}}, \ and\ \bibinfo {author} {\bibfnamefont {S.~T.}\ \bibnamefont {Sanders}},\ }\href {\doibase 10.1366/000370206776342625} {\bibfield  {journal} {\bibinfo  {journal} {Appl Spectrosc}\ }\textbf {\bibinfo {volume} {60}},\ \bibinfo {pages} {246} (\bibinfo {year} {2006})},\ \bibinfo {note} {publisher: SAGE Publications Ltd STM}\BibitemShut {NoStop}%
\bibitem [{ada()}]{adas_2024}%
  \BibitemOpen
  \href {https://www.adas.ac.uk/index.php} {}\bibinfo {howpublished} {\url{https://www.adas.ac.uk/index.php}}\BibitemShut {NoStop}%
\bibitem [{\citenamefont {Saleh}\ \emph {et~al.}(1998)\citenamefont {Saleh}, \citenamefont {Jost}, \citenamefont {Fei},\ and\ \citenamefont {Teich}}]{saleh1998entangled}%
  \BibitemOpen
  \bibfield  {author} {\bibinfo {author} {\bibfnamefont {B.~E.}\ \bibnamefont {Saleh}}, \bibinfo {author} {\bibfnamefont {B.~M.}\ \bibnamefont {Jost}}, \bibinfo {author} {\bibfnamefont {H.-B.}\ \bibnamefont {Fei}}, \ and\ \bibinfo {author} {\bibfnamefont {M.~C.}\ \bibnamefont {Teich}},\ }\href@noop {} {\bibfield  {journal} {\bibinfo  {journal} {Physical review letters}\ }\textbf {\bibinfo {volume} {80}},\ \bibinfo {pages} {3483} (\bibinfo {year} {1998})}\BibitemShut {NoStop}%
\bibitem [{\citenamefont {Burdick}\ \emph {et~al.}(2018)\citenamefont {Burdick}, \citenamefont {Varnavski}, \citenamefont {Molina}, \citenamefont {Upton}, \citenamefont {Zimmerman},\ and\ \citenamefont {Goodson~III}}]{burdick2018predicting}%
  \BibitemOpen
  \bibfield  {author} {\bibinfo {author} {\bibfnamefont {R.~K.}\ \bibnamefont {Burdick}}, \bibinfo {author} {\bibfnamefont {O.}~\bibnamefont {Varnavski}}, \bibinfo {author} {\bibfnamefont {A.}~\bibnamefont {Molina}}, \bibinfo {author} {\bibfnamefont {L.}~\bibnamefont {Upton}}, \bibinfo {author} {\bibfnamefont {P.}~\bibnamefont {Zimmerman}}, \ and\ \bibinfo {author} {\bibfnamefont {T.}~\bibnamefont {Goodson~III}},\ }\href@noop {} {\bibfield  {journal} {\bibinfo  {journal} {The Journal of Physical Chemistry A}\ }\textbf {\bibinfo {volume} {122}},\ \bibinfo {pages} {8198} (\bibinfo {year} {2018})}\BibitemShut {NoStop}%
\end{thebibliography}%

\end{document}